\documentclass[prc,aps,floats,floatfix]{revtex4}
\usepackage{graphicx}
\usepackage{color}
\usepackage{amsmath}
\usepackage{amssymb}
\usepackage{graphicx}
\usepackage{booktabs}
\usepackage{mathtools}
\usepackage{subfig}
\usepackage{multirow}
\usepackage{amsthm}
\usepackage{listings}
\usepackage{rotating}
\begin{document}

\title{Many-body approximations for atomic binding energies}

\author{Micah D. Schuster, Calvin W. Johnson, and Joshua T. Staker}
\affiliation{Department of Physics, San Diego State University,
5500 Campanile Drive, San Diego, CA 92182-1233}

\begin{abstract}
We benchmark three standard approximations for the many-body problem -- 
the Hartree-Fock, projected Hartree-Fock, and random phase 
approximations -- against full numerical configuration-interaction calculations of
the electronic structure of atoms, from Li through to Ne. These configuration-interaction 
calculations used up to $2 \times 10^8$ uncoupled basis states, equivalent to 
$ 10^7$ coupled basis states (configuration state functions.) 
Each method uses 
exactly the same input, i.e., the same single-particle basis and Coulomb matrix 
elements, so any differences are strictly due to the approximation itself. 
Although it consistently overestimates the ground state binding energy, 
the random phase approximation
has the smallest overall errors; furthermore, we suggest it may be useful as a method 
for efficient optimization of single-particle basis functions.
\end{abstract}
\maketitle

\section{Introduction}

In principle we know how to find the electronic structure and binding energies of atom: solve 
the many-body time-independent Schr\"odinger equation for $N$ electrons 
about a nucleus with charge $Z$:
\begin{equation}
\hat{H}\Psi = \left (\sum_{i = 1}^N -\frac{\hbar^2 }{2m} \nabla^2_i- \frac{Z e^2}{r_i} 
+ \sum_{i > j} \frac{e^2}{\left|\vec{r}_i - \vec{r}_j \right|} \right) 
\Psi = E \Psi
\label{Hcoord}
\end{equation}
or variants thereof including relativistic corrections. In practice one cannot solve 
this exactly, or even fully numerically much beyond helium.

So one turns to approximations. But for any approximation there are errors intrinsic to 
the approximation method itself, and errors due to the input, such as truncations in 
the space or assumptions in the Hamiltonian.  Surprisingly, rather little effort has been 
put into disentangling these two sources of error.

In this paper we benchmark three  approximations against an `exact' calculation with a fixed space and fixed nonrelativistic Hamiltonian.  For our exact calculation we use 
full configuration-interaction ( full CI); for a chosen single-particle basis, we can find numerically 
converged solutions for the ground state (as well as some number of excited states). 
The full CI spaces are large, up to dimension $10^8$ in the uncoupled (`M-scheme') basis, 
or $2 \times 10^6$ in the coupled basis of configuration state functions (CSFs);
see below for further discussion.
We also have a suite of codes that carry out Hartree-Fock, angular-momentum projected Hartree-Fock, and 
random phase approximations, using the identical input 
as CI. 
Thus our goal is to carefully document errors due the approximations themselves (and not due to 
the single-particle space, relativistic corrections or lack thereof, etc.); in this we are 
extending prior work benchmarking approximations for the nuclear many-body problem \cite{SJ02,SJ03,PHF}.


In the next section we define our many-body methods and discuss briefly computational 
issues. In Section 3 we give our results for several different spaces for second row atoms, from Li to 
Ne, comparing not only the ground state binding energies but also the experimentally more relevant 
ionization potentials and electron affinities. Because full CI is very time-consuming and our 
approximations computationally cheap, in Section 4 we consider the possibility of using 
approximations for fast optimization of the the single-particle basis, and of the considered 
methods find RPA the most accurate. 

\section{Many-body methods and inputs}

We work entirely in occupation or configuration space, meaning
we start with some set of orthonormal single-particle states, 
$\{ \phi_a(\vec{r}) \}$, with good angular momentum and parity, and introduce 
in second quantization \cite{LM85, BG77} fermion creation  and annihilation 
operators $\hat{a}^\dagger, \hat{a}$ that create or remove a particle from 
the states $ \phi(\vec{r})$. For definitiveness we have $N_s$ single-particle states. 
The Hamiltonian is then \cite{LM85,BG77,ring}
\begin{equation}
\hat{H}=\sum_{ab} H^1_{ab} \hat{a}^\dagger_a \hat{a}_b 
+ \frac{1}{4} \sum_{abcd} V_{ab,cd}\hat{a}^\dagger_a \hat{a}^\dagger_b \hat{a}_d \hat{a}_c 
\label{H2q}
\end{equation}
The one-body and two-body matrix elements, $H^{1}_{ab}$ (which includes both kinetic 
energy and the nuclear central potential) and $V_{ab,cd}$ respectively, 
are appropriate integrals over the Hamiltonian and the single-particle states.
Because the Hamiltonian is an angular momentum scalar, we generate  the matrix elements 
in coupled form. Given the single-particle space, the matrix elements are computed externally to 
all our codes and read in via a file. Any radial form of the single-particle states and any form of the 
interaction, including non-local interactions, are allowed and make no practical difference to our 
many-body codes.  Because we work in second quantization, the matrix elements are antisymmeterized and 
we do not separate direct from exchange terms. 

For this paper we restrict the Hamiltonian to the simple nonrelativistic Coulomb interaction (\ref{Hcoord}).
We build our single-particle radial wavefunctions from Slater-type orbitals (STOs), that is, 
functions of the form $r^n \exp(-\beta r)$, which allows us to compare to other work as well as making 
the integrals analytic.  We orthonormalize our single-particle basis so that the one-body part of 
(\ref{H2q}) is diagonal. 
For our specific choices of parameters for our basis, we use the CVB$n$ bases sets \cite{Ema03}, 
described below. 
Finally, we describe the atom as a two-species systems, with spin-up and spin-down electron occupying
distinct orbits.  We fix $M_s = (n_\uparrow - n_\downarrow)/2$ in all calculations, as 
we have no spin-orbit coupling.

These methods were chosen because we have existing codes for them, and thus leave for future work 
other methods such as the multi-configurational Hartree-Fock approximation \cite{FFBJ97,Cook98} 
and many-body 
perturbation theory \cite{LM85,Je07,Cook98}

\subsection{Configuration-interaction calculations}

For full configuration-interaction (full CI) calculations \cite{BG77,Lo55CI,CIV3,GRASP92, SS99,Cook98,Je07,BW88,Sh98},
 we use the BIGSTICK code \cite{BIGSTICK}, developed originally for 
nuclear CI calculations  but which can be applied as well to atomic calculations. The only assumption is that the 
single-particle basis be states of good angular momentum and parity. 

BIGSTICK creates a many-body basis of Slater determinants $\{ | \alpha \rangle \}$, constructed 
from the single-particle basis:
\begin{equation}
| \alpha \rangle = \prod_{i = 1}^N \hat{a}^\dagger_i | 0 \rangle,
\end{equation}
where $N$ is the number of particles (occupied states).  
Furthermore, because we have no spin-orbit coupling (a generalization we leave to future work), we fix 
the number of spin-up and spin-down electrons and thus fix $M_s = (n_\uparrow -n_\downarrow)/2$. 
Because  single-particle states each have good $l_z$, 
it is almost trivial to generate many-body basis states with fixed total $M = L_z$; in nuclear 
CI calculations this is called an $M$-scheme basis. And because the Hamiltonian is an angular momentum scalar, 
the final eigenstates thus will automatically have good total $L$ (and $S$). Although we have the capability 
for various truncations on the many-body basis, such as allowing only single- or double-electron excitations, 
we make no such truncations here as they have no correspondence for our approximation methods. 

Most atomic CI programs such as CIV3 \cite{CIV3} and 
GRASP92 \cite{GRASP92} use a basis of configuration state functions (CSFs) coupled up to good 
$L$ and $S$ (or total $J$, hence in nuclear structure physics this is called a $J$-scheme basis). 
Such a basis is smaller in dimension, by a factor of 10 to 100, but there are always trade-offs: the basis states are linear 
combinations of $M$-scheme Slater determinants and hence the many-body Hamiltonian matrix elements are 
more complicated to calculate. And although the dimensions of an uncoupled $M$-scheme basis are larger, 
the Hamiltonian matrix is much sparser.   Furthermore the simplicity of the uncoupled basis allows us to avoid 
storing the many-body matrix element, and instead recompute them 
efficiently on the fly \cite{ANTOINE}, which further reduces 
the storage of the Hamiltonian by a factor of 10 to 100. 
We can go up to uncoupled basis dimensions of $10^8$ on a desktop machine (and similar programs on parallel 
computers reach up to $10^{10}$), as listed in Table II, while the dimension of the largest equivalent 
coupled CSF ground state dimension is about $ 10^7$, as listed in Table III.

Given these arguments, it is interesting to compare basis sizes. These  are not always given in 
the literature, and many  `large-scale' calculations rely upon careful pruning 
of the configuration basis \cite{Correge04,Gupta05,Deb09}; nevertheless, for comparison 
purposes a  large scale relativistic CI calculation \cite{Chen01} 
used up to 197,426 basis CSFs, while a more recent nonrelativistic calculation \cite{Bro10} used up to 
901,816 CSFs. Quantum chemists surpassed the one-billion determinant limit some time ago \cite{Sh98,OJS90}.

In a given uncoupled basis, BIGSTICK computes the many-body Hamiltonian matrix elements 
$\langle \alpha | \hat{H} | \beta \rangle$ and finds the low-lying eigenstates, including the ground state, using the Lanczos algorithm \cite{Lanczos} (the Davidson-Liu algorithm \cite{Dav93,LSAS01} is efficient for the 
diagonal-dominated atomic problem but not the more general many-body problem such as nuclear CI; furthermore recomputing the matrix elements on the fly is better suited for the Lanczos algorithm). 
Convergence of, say, five states take a few hundred Lanczos iterations; 
on a desktop machine with eight OpenMP threads 
we can accomplish this for an $M$-scheme basis in a few hours or at most a few days.

With the single-particle space and Hamiltonian matrix elements fixed, the resulting eigenenergies for the 
full-space CI calculation are numerically exact. All other calculations described in this paper are 
numerical approximations to these results. 

CI calculations have been previously carried out for second row elements, see for example 
nonrelativistic \cite{Weiss61,Bu68,Saski74,Bunge76,Silver79,Jitrik97,Bunge06}  and relativistic calculations \cite{Chen01,Almora10} , though 
most of these were not full CI and thus were in much smaller dimensions.

\subsection{Hartree-Fock approximation}

The Hartree-Fock approximation (HF) is a variational method using a single Slater determinant \cite{Lo55HF,FFBJ97,Cook98,Je07}. In occupation space \cite{ring} one applies 
a unitary transformation among the single-particle states:
\begin{equation}
\hat{c}_i^\dagger = \sum_{a=1}^{N_s} U_{ia} \hat{a}_a^\dagger, \label{Usp}
\end{equation}
and creates the trial Slater determinant 
\begin{equation}
| \Psi_T \rangle =\prod_{i =1}^N \hat{c}^\dagger_i | 0 \rangle.
\end{equation}
Then one finds a Slater determinant that minimizes the energy, that is, that minimizes
\begin{equation}
E = \frac{ \langle \Psi_T | \hat{H} | \Psi_T \rangle } { \langle \Psi_T |  \Psi_T \rangle }
\end{equation}

Because our input matrix elements are  antisymmeterized, we fully include both direct and exchange 
terms.  Working in occupation space the exchange term causes us no difficulty (or, to put it another way, 
any difficulty is off-shored into calculation of the antisymmeterized integrals). 

Many Hartree-Fock calculations enforce good angular momentum, for closed-shell systems or 
closed-shell plus or minus one particle.  Our HF code allows the Slater determinant to break rotational 
invariance, even for closed-shell systems; such Slater determinants are `deformed.'  
Experience in nuclear physics suggest this to be a fruitful 
path. In a few cases, when possible, we will compare and contrast spherical and deformed HF solutions. 

In order to minimize, we use the standard Hartree-Fock equations, which consist of 
iteratively solving
\begin{equation}
\mathbf{h}\vec{u}_i = \epsilon_i \vec{u}_i, \label{HF}
\end{equation}
where the Hartree-Fock effective one-body Hamiltonian is 
\begin{equation}
h_{ab} = T_{ab} + \sum_{cd} V_{ac,bd} \rho_{dc}
\end{equation}
and 
\begin{equation}
\rho_{dc} = \sum_{i = 1}^N U_{ic} U_{id}.
\end{equation}
Internally we use an $N_s \times N$ rectangular matrix $\mathbf{\Psi}$, with the $N$
columns representing the transformed occupied states $\hat{c}^\dagger_i$. In that 
case $\mathbf{\rho} = \mathbf{\Psi \Psi}^\dagger$.  

(Because we leave out spin-orbit coupling and conserve $n_\uparrow$ and $n_\downarrow$, 
we use  separate Slater determinants for spin-up and spin-down electons; this is a 
straightforward generalization. )

The Hartree-Fock energy is now
\begin{equation}
E_\mathrm{HF} = \sum_{i=1}^N \left( \epsilon_i -\frac{1}{2} \sum_{cd} 
V_{ic,id} \rho_{dc} \right),
\end{equation}
where we have to subtract off the potential to keep from double-counting in the 
single-particle energies.

\subsection{Projected Hartree-Fock approximation}

If one does not have a closed shell, a single Slater determinant will not in general 
have good angular momentum.  Conversely, an open-shell state with good angular momentum,
which can be written as a sum of Slater determinants, will have important correlations that 
lower the energy.

One can start with single-particle states which have good angular momentum and then construct 
many-body states with good total angular momentum; these are called configuration state functions (CSFs), 
and have been applied to second-row elements, e.g. \cite{ Lask75,Lunell68}.

We take a different approach, however. Our Hartree-Fock code allows for arbitrary deformation: 
the single-particle states generated by the self-consistent field do not have good angular momentum. 
The Slater determinant built from these states is called a `deformed' state. 
In order to project out good angular momentum, we introduce the 
standard projection operator \cite{La80,ring}
\begin{equation}
\hat{P}^J_{MM^\prime} = 
\int d\Omega {\cal D}^{(J)*}_{MM^\prime} (\Omega) \hat{R}(\Omega),
\label{jproj0}
\end{equation}
where $\Omega =$ the Euler angles $ \alpha,\beta,\gamma$ and $d\Omega = d\alpha \sin \beta d\beta d\gamma$,
${\cal D}^{(J)*}_{MM^\prime}$ is the Wigner $D$-matrix \cite{Edmonds}, and 
$\hat{R}(\Omega)$ is the rotation operator.  
To compute 
$\langle \Psi | \hat{R}(\Omega) | \psi \rangle$ and 
$\langle \Psi | \hat{H} \hat{R}(\Omega) | \psi \rangle$, we use the 
matrix representation 
\begin{equation}
\hat{R}| \Psi \rangle \rightarrow \mathbf{R} \mathbf{\Psi} = \tilde{\mathbf{\Psi}}
\end{equation}
where $\mathbf{\Psi}$ and $\tilde{\mathbf{\Psi}}$ are $N_s \times N$ matrices 
representing our original and rotated Slater determinants, respectively, and 
$\mathbf{R}$ is a square $N_s \times N_s$ matrix;  in our single-particle 
basis the matrix elements are straightforward:
\begin{equation}
R_{ab} = \langle j_a m_a | \hat{R}(\alpha, \beta, \gamma) | j_b m_m \rangle
= \delta_{j_a j_b} {\cal D}^{j_a}_{m_a, m_b}(\alpha, \beta, \gamma),
\end{equation}
where ${\cal D}^{j}_{m^\prime, m}$ is the Wigner $D$-matrix \cite{Edmonds}.
It is useful to note that the $j,m,m^\prime$ for the rotational matrix $\hat{R}$ are those 
of the single-particle space, while the $J,M, M^\prime$ for the projection operator 
(\ref{jproj0}) are those of the many-body space. 

Matrix elements between two oblique Slater determinants is 
relatively straightforward \cite{Lang93}. First, 
\begin{equation}
\langle \Psi | \hat{R}|{\Psi} \rangle =
\langle \Psi | \tilde{\Psi} \rangle = \det \Psi^\dagger \tilde{\Psi}.
\end{equation}
Calculation of the Hamiltonian (and other) expectation value requires 
the density matrix, 
\begin{equation}
\rho^\prime_{ab} = \frac{ \langle \Psi | \hat{\phi}^\dagger_a 
\hat{\phi}_b|\tilde{\Psi} \rangle }
{ \langle \Psi | \tilde{\Psi} \rangle} 
= \left [ 
\tilde{\mathbf{\Psi}} \left(  \mathbf{\Psi}^\dagger \tilde{\mathbf{\Psi}}\right )^{-1} 
\mathbf{\Psi}^\dagger \right ]_{ba}
\end{equation}
Then 
\begin{equation}
\langle \Psi | \hat{H} | \tilde{\Psi} \rangle 
= \sum_{ab} H^1_{ab} \rho^\prime_{ab} 
+ \frac{1}{2}\sum_{abcd}V_{ab,cd}( \rho^\prime_{ac} \rho^\prime_{bd} - \rho^\prime_{ad} \rho^\prime_{bc} )
\end{equation}
Accounting for two species is straightforward.
Note that even if our original Slater determinants were real, by rotating over all 
angle they can become complex. 

We only project out good orbital angular momentum $L$. We could in principle project out 
exact spin $S$; we leave such a modification for 
future work. Because we leave off spin-orbit coupling, however, we automatically get states 
with good $S$ already.

We now can compute the Hamiltonian and ``norm'' matrix elements,
\begin{equation}
h^J_{M,M^\prime} \equiv \langle \Psi|
  \hat{H} \hat{P}^J_{M,M^{\prime}}   | \Psi \rangle , \,\,\,
n^J_{M,M^\prime} \equiv  \langle \Psi|
  \hat{P}^J_{M,M^{\prime}}   | \Psi \rangle,
\end{equation}
which are both Hermitian.
 Then we solve the
generalized eigenvalue equation
\begin{equation} 
\sum_{M^\prime} h^J_{M,M^\prime} g^J_{M^\prime} = E^J_{PHF} \sum_{M^\prime} n^J_{M,M^\prime} g^J_{M^\prime}.
\end{equation}
where we  allow $g_M$ to be complex.  Although the deformed 
Hartree-Fock state will have an orientation, the final result will be independent of 
orientation; we confirmed this by arbitrarily rotating our HF state.

Both the Hartree-Fock and the projected Hartree-Fock energies will be variational 
upper bounds to the exact CI energy, with the projected Hartree-Fock energy lower than 
(or equal to, in the HF state has good rotational symmetry) $E_{HF}$. 

\subsection{Random phase approximation}

Once one has a Hartree-Fock calculation, it is straightforward to go on to the random 
phase approximation, using the matrix formulation \cite{ring,SJ02}:
\begin{equation}
 \left(
\begin{array}{cc}
\mathbf{A} & \mathbf{B} \\
-\mathbf{B}^* & -\mathbf{A}^*
\end{array}
\right ) \left( \begin{array}{c} \vec{X}_\lambda \\
\vec{Y}_\lambda
\end{array} \right )
= \hbar \Omega_\lambda \left( \begin{array}{c} \vec{X}_\lambda \\
\vec{Y}_\lambda
\end{array} \right )
\end{equation}
where
\begin{equation}
A_{nj,mi} \equiv \left \langle  \left [\left [ \hat{c}^\dagger_j \hat{c}_n, \hat{H} \right ], \hat{c}^\dagger_m
\hat{c}_i \right ]  \right \rangle
\end{equation}
and
\begin{equation}
B_{nj,mi} \equiv \left \langle  \left [ \hat{c}^\dagger_i
\hat{c}_m ,\left  [ \hat{c}^\dagger_j \hat{c}_n, \hat{H} \right ] \right ]  \right \rangle.
\end{equation}
Here $i,j$ label occupied single-particle levels in the Hartree-Fock state, 
while $m,n$ label unoccupied levels. 

From this we can compute the RPA correlation energy, which is a correction to the
HF energy.  When working with a rotationally invariant Hamiltonian, as we do here, 
it is possible to have a HF state which breaks rotational symmetry; this means one 
could rotate the HF state in any direction and not change the HF energy. In RPA 
this shows up as a zero-energy mode, which must be dealt with carefully 
when computing the binding energy \cite{ring,SJ02}.  The final result is
\begin{equation}
E_\mathrm{RPA} = E_{HF} -\frac{1}{2}{\rm Tr}(A) + \frac{1}{2}\sum_{i}\hbar \Omega_i.
\end{equation}

The RPA energy, while lower than the HF energy, is not variational. In previous 
calculations for nuclei in valence spaces \cite{SJ02} RPA both under- and overestimated 
the binding energy; in our chosen atomic cases below, RPA consistently overestimates the 
binding energy.  Because we start from a deformed HF state, our RPA calculations do 
not have good angular momentum. (RPA is often said to `restore' rotational symmetry, 
but this is not restoration of good angular momentum quantum numbers. Rather, 
the Hartree-Fock energy is invariant under rotation, and this invariance shows up as 
a $\Omega=0$ mode that appears in RPA but not in the simpler Tamm-Dancoff approximation 
or TDA \cite{ring}. In this sense RPA `restores' rotational invariance that was broken in TDA.)

\subsection{Input spaces and interaction}

Our codes can use (up to memory limitations, of course) any finite set of 
single-particle states which have good angular momentum. For convenience we 
use the CVB1, CVB2, and CVB3 Slater-type orbitals (STOs) and parameter values of Ema 
\textit{et al.} \cite{Ema03}. These are radial wavefunctions of the 
form $r^n \exp(-\beta r)$, with the set of STOs and the values of $\beta$ optimized 
for Li through Ne. The single-particle spaces ranged from 8 orbits 
(for CVB1-Li) up to 22 (for CVB3-Ne).

\begin{table}[h]
	\centering
\caption[Basis Set Composition]{Basis Set Composition \label{table:basislist}}
\begin{tabular}{|c|ccccc|cccc|}
\hline
\multicolumn{2}{|c}{~~}&~~&\multicolumn{3}{c}{Li and Be}&~~&\multicolumn{3}{c|}{B to Ne}\\
Orbit&&&CVB1&CVB2&CVB3&&CVB1&CVB2&CVB3\\
\hline
s&&&6&7&8&&6&7&8\\
p&&&2&3&4&&4&5&6\\
d&&&0&2&3&&1&3&4\\
f&&&0&0&2&&0&1&3\\
g&&&0&0&0&&0&0&1\\
\hline
Total&&&8&12&17&&11&16&22\\
\hline
\end{tabular}
\end{table}

To construct our many-electron basis states, we fixed $n_\uparrow$ spin-up 
electrons and $n_\downarrow$ spin-down electrons, with $n_\uparrow = n_\downarrow$ 
or $=n_\downarrow+1$, which allowed us to regain states of all possible total $S$. We could and did choose 
$n_\uparrow + n_\downarrow \neq Z$ in order to compute ionization potentials and 
electron affinities. We also fixed total $M = L_z = 0$, which allowed us to 
regain states of all possible total $L$.  We otherwise had no constraints, e.g., we did not 
restrict ourselves to single, double, or otherwise excitations; although our 
CI code can make such restrictions, the HF, PHF, and RPA cannot.  Dimensions 
of the uncoupled many-body CI spaces we used are in Table \ref{table:cispace}, while the 
equivalent dimensions for coupled (CSF) bases for the ground states are in Table 
\ref{table:cispacecoupled}, where we indicated the ground state quantum numbers 
by $^{2S+1}(L)^\pi$.

\begin{table}[h!]
\centering
\caption[Size of the CI Space]{Dimensions of the uncoupled ($M$-scheme) CI spaces \label{table:cispace}}
\begin{tabular}{|l|rrrr|}
\hline
Atom &$~$& \multicolumn{3}{c|}{Basis Set}\\
&&\multicolumn{1}{c}{CVB1} & \multicolumn{1}{c}{CVB2} & \multicolumn{1}{c|}{CVB3} \\
\hline
Li && $192$ & $1067$ & $4958$ \\
Be && $866$ & $10983$ & $100842$ \\
B  && $46116$ & $909534$ & $1.1\times10^7$ \\
C  && $300309$ & $1.1\times10^7$ & $2.5 \times 10^8$ \\
N  && $1.4\times10^6$ & $1.1\times10^8$ & \multicolumn{1}{c|}{--}\\
O  && $6.6\times10^6$ &\multicolumn{1}{c}{--} & \multicolumn{1}{c|}{--}\\
F  && $2.4\times10^7$ &\multicolumn{1}{c}{--} & \multicolumn{1}{c|}{--} \\
Ne && $8.7\times10^7$ &\multicolumn{1}{c}{--} & \multicolumn{1}{c|}{--} \\
\hline
\end{tabular}
\end{table}

\begin{table}[h!]
\centering
\caption[Dimensions of the Coupled CI Space]{Size of the equivalent coupled (CSF or $J$-scheme) CI spaces \label{table:cispacecoupled}}
\begin{tabular}{|l|c|rrrr|}
\hline
Atom & $^{2S+1}(L)^\pi$  &$~$& \multicolumn{3}{c|}{Basis Set}\\
& (g.s.)&& \multicolumn{1}{c}{CVB1} & \multicolumn{1}{c}{CVB2} & \multicolumn{1}{c|}{CVB3} \\
\hline
Li & $^2 S^+$ && $94$ & $223$ & $516$ \\
Be & $^1 S^+$  && $190$ & $775$ & $3102$ \\
B  & $^2 P^-$ && $7752$ & $82429$ & $618245$ \\
C  & $^3 P^+$ && $35126$ & $770145$ & $1.1 \times 10^7$  \\
N  & $^4 S^-$ && $53710 $ & $2.1\times 10^6 $ & \multicolumn{1}{c|}{--}\\
O  & $^3 P^+$ && $604146$ &\multicolumn{1}{c}{--} & \multicolumn{1}{c|}{--}\\
F  & $^2 P^-$ && $1.8 \times 10^6$ &\multicolumn{1}{c}{--} & \multicolumn{1}{c|}{--} \\
Ne & $^1 S^+$ && $1.3 \times 10^6$ &\multicolumn{1}{c}{--} & \multicolumn{1}{c|}{--} \\
\hline
\end{tabular}
\end{table}

To construct our interaction file, we first orthornormalized the STO basis. 
We did this by diagonalizing the one-body part of the Hamiltonian (and so this 
depended on the $Z$ charge of the nucleus); then the one-body Hamiltonian is 
diagonal and expressed in terms of single-particle energies.  The two-body matrix 
elements were first computed with the STOs, which can be done semi-analytically, and 
then transformed to the orthonormalized basis.  Our matrix elements and resulting 
atomic energies for small cases were validated against an prior, independent code (M. Bromley, 
private communication).

\section{Results}

The following sections give descriptions of the results obtained as well as tables and example figures. The computed ground state energies for HF, PHF, and RPA are compared to CI. The ionization potentials and electron affinities are compared to experimental values 
 \cite{lide}.

\subsection{Ground State Energy}
The computed ground state energies for the neutral atoms studied are in Table \ref{tab:gs}.  CI is exact for purposes of comparison. As required by variational theory, Hartree-Fock gives an upper bound to the ground state energy and projected Hartree-Fock improves upon it. The random phase approximation computes lower energies than CI in all of the calculations we performed. There are no CI calculations for several the largest basis sets in combination with the largest atoms because the computational requirements exceeded our available resources; keep in mind that our `approximate' results were computed in 
full CI spaces, as we allow arbitrary deformation of the single-particle states. 

\begin{figure}
\centering
\includegraphics[width=0.6\textwidth, trim=0mm 0mm 0mm -25mm]{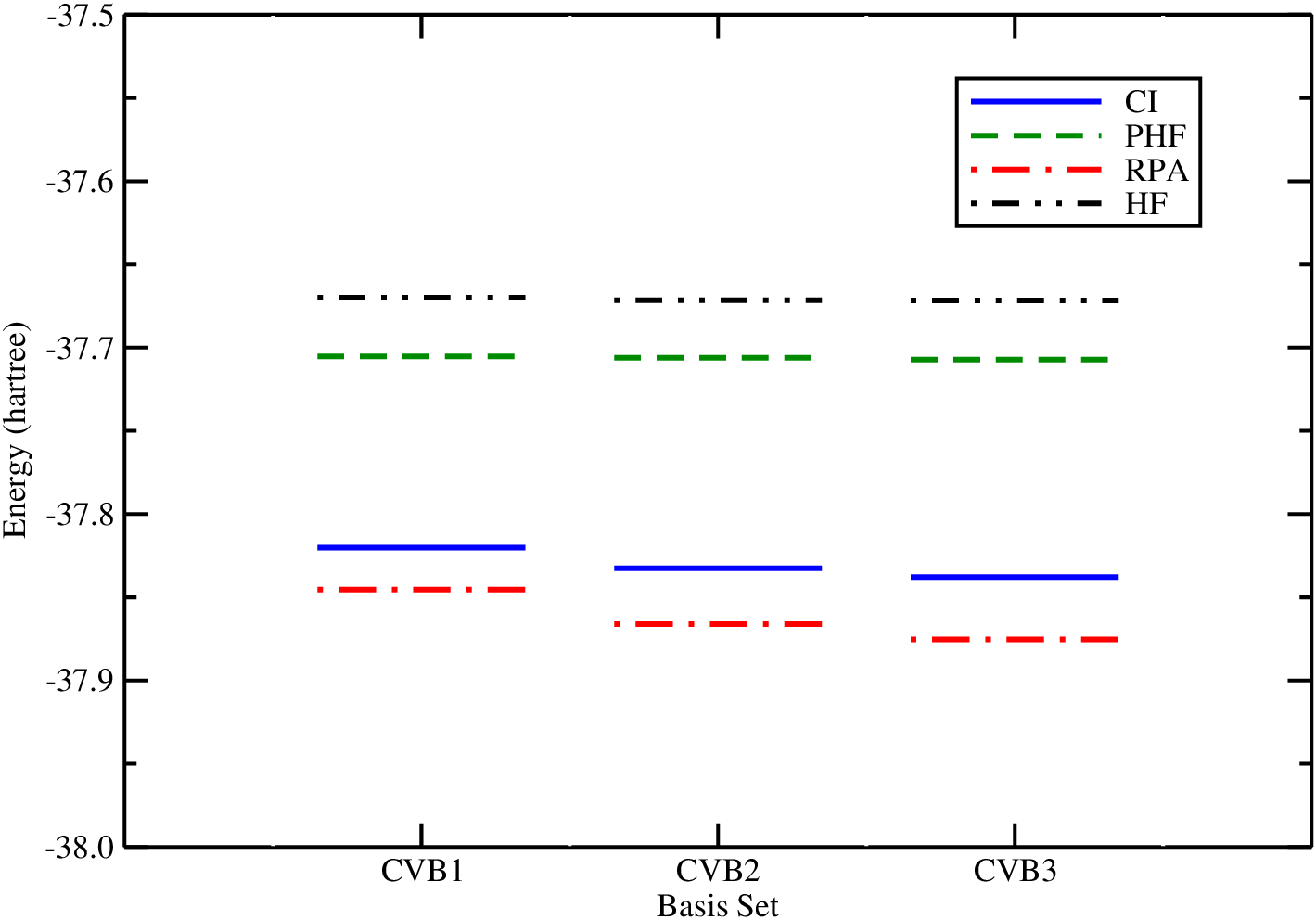}   
\caption{(Color online.) Carbon ground state energy. \label{fig:carbongs}}
\end{figure}

Figure \ref{fig:carbongs} shows the results for carbon in graphical form; the results for all other 
atoms are qualitatively similar, although the error for all approximations increases as we add more 
electrons. For example, the error in RPA is only a few millihartrees for Li, while it is 100 millihartrees 
for Ne. 

Note there is no difference between the HF and PHF results for Li and for Ne; this is because 
the HF states for neutral
Li and Ne are already in states of good total L; projection cannot change this.

\begin{table}

\caption[Ground State Energy]{Ground State Energy (hartree)\label{tab:gs}}
\begin{tabular}{lr|rrrrrrr}
\hline
\multicolumn{1}{c}{Atom} & \multicolumn{1}{c}{space} & \multicolumn{1}{c}{HF} &  & \multicolumn{1}{c}{PHF} & &\multicolumn{1}{c}{RPA} & & \multicolumn{1}{c}{CI} \\
\hline

    & CVB1 &-7.4327  & &-7.4327& &-7.4720  & &-7.4700   \\ 

Li  & CVB2 &-7.4327  & &-7.4327& &-7.4751  & & -7.4725  \\ 

    & CVB3 &-7.4327  & &-7.4327& &-7.4774  & &-7.4741  \\ 
\hline

  & CVB1 &-14.5732& &-14.5758& &-14.7491& &-14.6558 \\

Be  &CVB2  &-14.5733& &-14.5778& &-14.7271& &-14.6589 \\

  & CVB3 &-14.5733& &-14.5780& &-14.7315& &-14.6615 \\
\hline

   & CVB1 &-24.5239& &\multicolumn{1}{c}{-24.5352}& &-24.6852 & &-24.6391 \\

B   & CVB2  &-24.5330& &-24.5353& &-24.6965& &-24.6458 \\

   & CVB3 &-24.5331& &-24.5355& &-24.7009& &-24.6482 \\
\hline

   & CVB1 &-37.6700 & &-37.7052 & &-37.8453& &-37.8202  \\

C   & CVB2 & -37.6715 & &-37.7061& &-37.8661& &-37.8326  \\

   & CVB3  &-37.6717& &-37.7072& &-37.8753& &-37.8378 \\

\hline

   & CVB1 &-54.3443& &-54.4221& &-54.5671& &-54.5517 \\

N   & CVB2  &-54.3445& &-54.4211& &-54.5971& &-54.5699      \\

   & CVB3 &-54.3473& &-54.4236& &-54.6121& &\multicolumn{1}{c}{--} \\

\hline

   & CVB1 &-74.7793& &-74.8267& &-75.0487& &-75.0037   \\

O   & CVB2 &-74.7822& &-74.8303& &-75.0995& &\multicolumn{1}{c}{--}  \\

   & CVB3 &-74.7829& &-74.8309& &-75.1172& &\multicolumn{1}{c}{--}  \\

\hline

   & CVB1  &-99.4133& &-99.4156& &-99.7283 & &-99.6457   \\

F   & CVB2 &-99.4155& &-99.4188& &-99.7943 & &\multicolumn{1}{c}{--}  \\

   & CVB3 &-99.4162& &-99.4195& &-99.8184 &  &\multicolumn{1}{c}{--}  \\

\hline

  & CVB1 &-128.5455& &-128.5455& &-128.9307& &-128.8278  \\

Ne  & CVB2  &-128.5462& &-128.5463& &-129.0083& &\multicolumn{1}{c}{--}      \\

  & CVB3 &-128.5470& &-128.5470& &-129.0399&      &\multicolumn{1}{c}{--}  \\

\hline
\end{tabular}

\end{table}

\subsection{Ionization Potential}
\begin{figure}
\centering
\includegraphics[width=0.6\textwidth, trim=0mm 0mm 0mm -25mm]{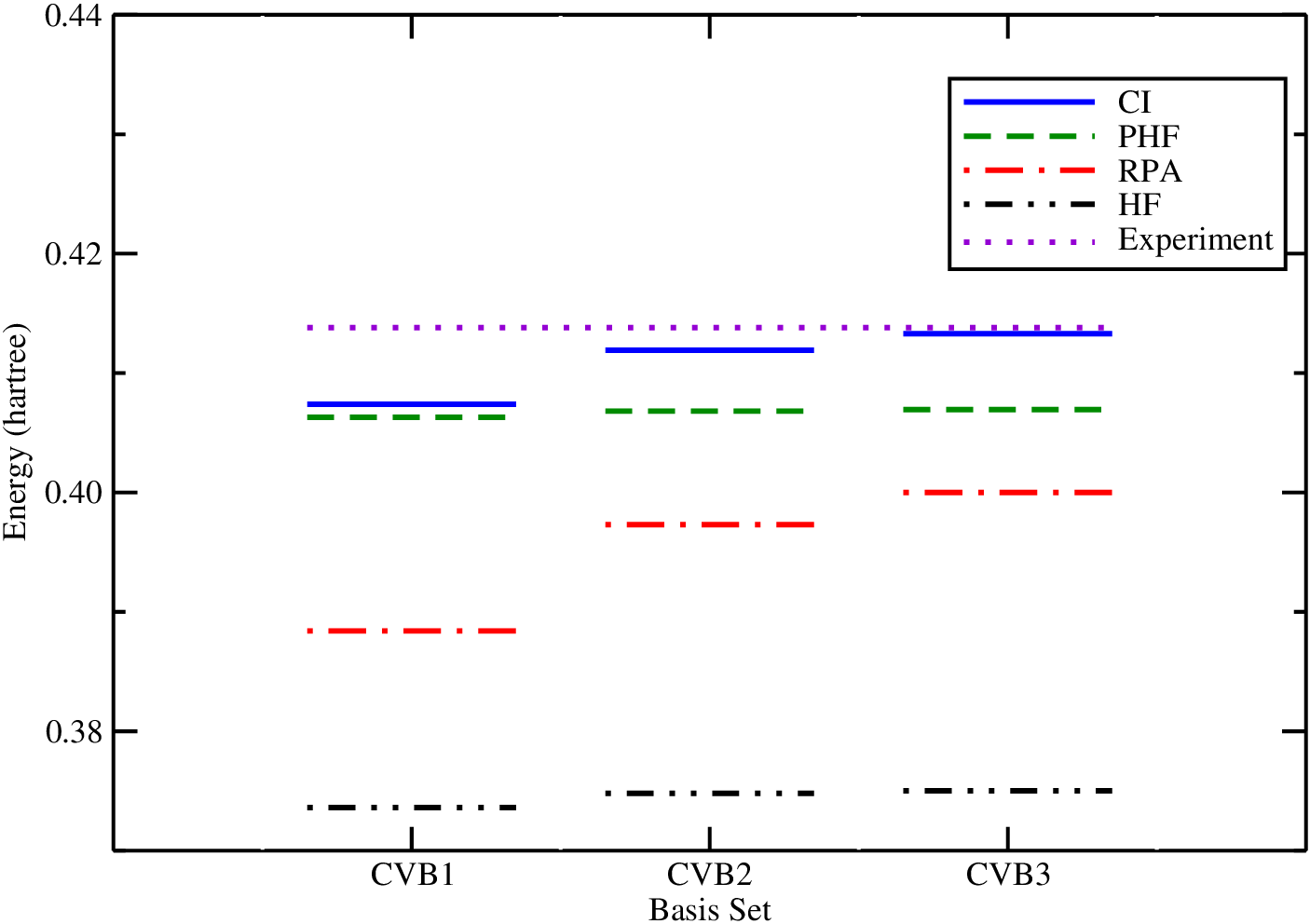}
\caption{(Color online.) Carbon ionization potential. \label{fig:carbonion}}
\end{figure}

Table \ref{tab:ion} shows the single ionization potentials for all methods as well as experimental values \cite{lide} for all atoms studied. Figure \ref{fig:carbonion} shows the ionization potential for carbon graphically.

Because the ionization potential (and in the next section, the electron affinity) is the difference between two ground state energies, the nice clean trends of the previous section are lost. 
CI agrees well with experiment, within a few millihartree through C and less than 20 millihartree difference
through Ne. Increasing basis size brings better agreement with experiment, though of course we note 
we have left out relativistic corrections. 

HF consistently underestimates the ionization potentials, with PHF  closer for Li through N (with 
HF and PHF identical for Li as the neutral Li and Li I HF states both have good angular momenta) but 
farther away for O, F, and Ne.

RPA results are the sporadic, sometimes better than HF / PHF and sometimes; again, this is because 
errors in the approximation do not cancel but exacerbate. Increasing the basis size does not always
lead to better estimates; they get worse for O, F, and Ne.

\begin{table}

\caption[Ionization Potential]{Ionization Potential (hartree)\label{tab:ion}}

\begin{tabular}{lr|rrrrrrrr|r}
\hline
\multicolumn{1}{c}{Atom} & \multicolumn{1}{c}{space} & \multicolumn{1}{c}{HF} &  & \multicolumn{1}{c}{PHF} & &\multicolumn{1}{c}{RPA} & & \multicolumn{1}{c}{CI} & & \multicolumn{1}{c}{Exp \cite{lide}}\\
\hline

  & CVB1  &0.1963& &0.1963& &0.1972& &0.1969 & & \\ 

Li  &CVB2  &0.1963& &\multicolumn{1}{c}{0.1963}& &0.1974& &0.1971& & 0.1982\\ 

  & CVB3 &0.1963& &0.1963& &0.1982& &0.1978& &\\ 

\hline

  & CVB1 &0.2957& &0.2993& &0.4324& &0.3407& & \\

Be  & CVB2 &0.2958& &0.3004& &0.4072& &0.3411& & 0.3427\\

  & CVB3 &0.2958& &0.3006& &0.4090& &0.3416& & \\

\hline
   & CVB1 &0.2955& &\multicolumn{1}{c}{0.2978}& &0.1908& &0.3002& & \\

B   & CVB2 &0.2955& &0.2978& &0.1949& &0.3030& & 0.3051\\

   & CVB3 &0.2955& &0.2979& &0.1974& &0.3041& & \\

\hline

   & CVB1 &0.3736& &0.4063& &0.3884& &0.4074  & & \\

C   & CVB2 &0.3748& &0.4068& &0.3973& &0.4119    & & 0.4140\\

   & CVB3 &0.3750& &0.4074& &0.4000& & 0.4133    & & \\

\hline

   & CVB1 &0.4860& &0.5166& &0.5218& &0.5259    & & \\

N   & CVB2  &0.4867& &0.5178& &0.5312& &0.5325  & & 0.5383\\

   & CVB3 &0.4870& &0.5168& &0.5343& &\multicolumn{1}{c}{--}    & & \\

\hline

   & CVB1 &0.4857& &0.4320& &0.5201& &0.4787   & & \\

O   &CVB2  &0.4852& &0.4335& &0.5350& &\multicolumn{1}{c}{--}    & & 0.5007\\

   & CVB3 &0.4859& &0.4340& &0.5408&     &\multicolumn{1}{c}{--}    & & \\

\hline

   &CVB1  &0.6241& &0.5674& &0.6670& &0.6216& & \\

F   &CVB2  &0.6215& &0.5652& &0.6792&    &\multicolumn{1}{c}{--}    & & 0.6405\\

   & CVB3 &0.6220& &0.5674& &0.6858&    &\multicolumn{1}{c}{--}    & & \\

\hline

  & CVB1 &0.7258& &0.7235& &0.8031& &0.7769   & & \\

Ne  & CVB2 &0.7220& &0.7186& &0.8128& &\multicolumn{1}{c}{--}    & & 0.7928\\

  & CVB3 &0.7227& &0.7192& &0.8206&   &\multicolumn{1}{c}{--}    & & \\

\hline
\end{tabular}
\end{table}

\subsection{Electron Affinity}

\begin{figure}[h]
\centering
\includegraphics[width=0.6\textwidth, trim=0mm 0mm 0mm -25mm]{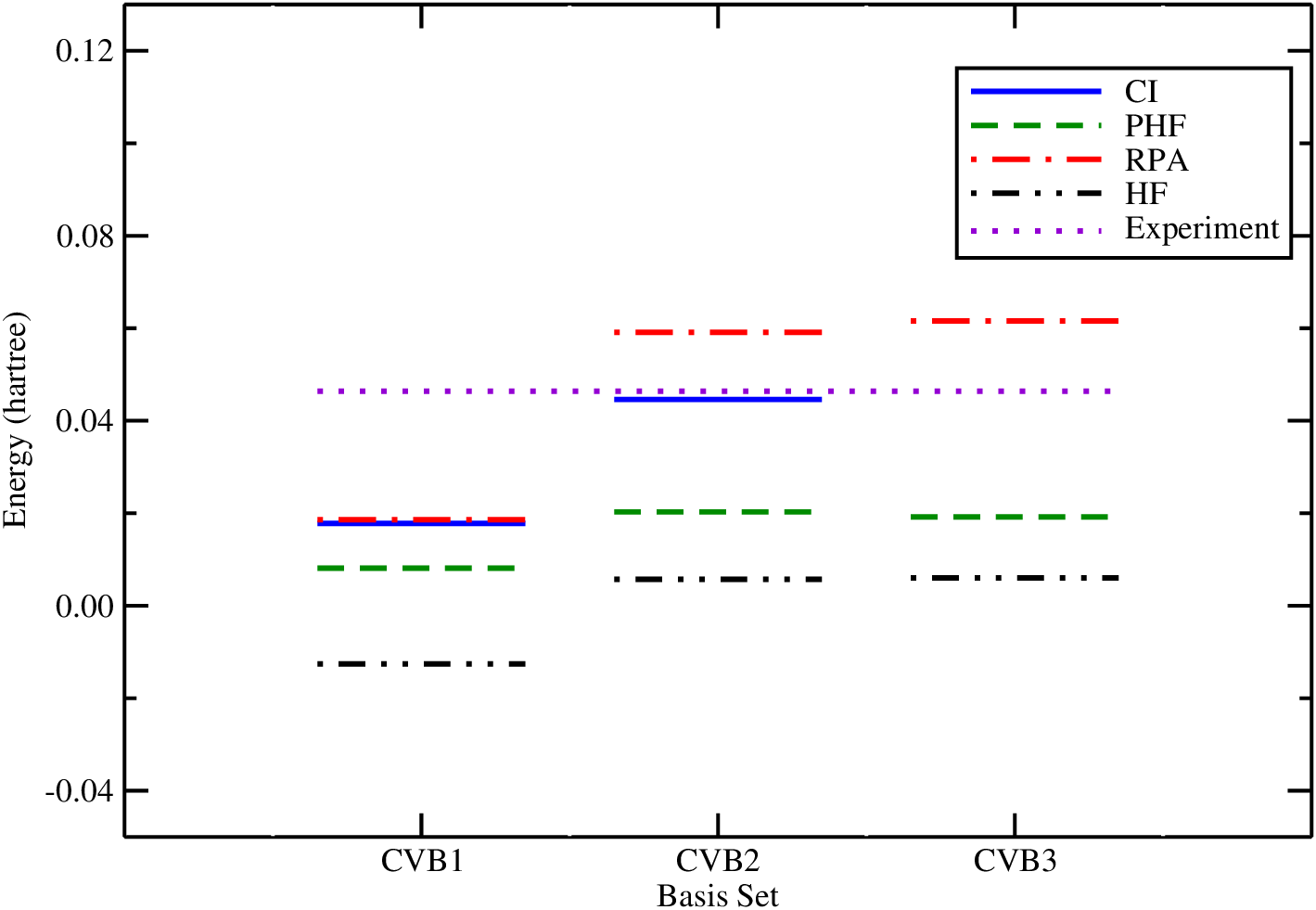}
\caption{(Color online.) Carbon electron affinity. There are no CI calculation for the  CBV3 
sets because the computational requirements exceeded our available resources. \label{fig:carbonaffin}}
\end{figure}

Not all of the second row atoms can bind an additional electron; we only present those cases which 
do so experimentally:   lithium, boron, carbon, oxygen and fluorine. Table \ref{tab:affinity} shows the electron affinities computed by each method and the respective experimental values \cite{lide}. Our results are comparable to other work in the field \cite{Cole82}. Figure \ref{fig:carbonaffin} shows the electron affinity of carbon in graphical form.

The quality of agreement is generally poor, even for CI.  In retrospect this is not surprising: 
we have already seen errors for all methods grow rapidly with the number of electrons, and given the 
smallness of second row electron affinities the results are unsurprising if disappointing. Many of 
the affinities even have the wrong sign.  Adding an electron to the atom increases the runtimes and memory requirements over the neutral atom calculations, therefore, only the smallest basis sets for all but lithium could be used. Without the larger calculations, a clear trend is difficult to establish. It appears that the CVB1 basis set is fairly poor when computing electron affinities, sometimes giving negative results. However, based on the CVB2 calculations for boron and carbon, the CI electron affinities might converge on the experimental values as basis size increases, similar to what is seen with the ground state energy and ionization potential calculations.

All three of our approximations perform poorly, trending neither towards the CI result nor towards 
experiment.  HF and PHF are unable to provide consistent results across either atoms or basis sets and generally tend to be too far from experimental values to be useful; nor do they show consistent trends as the basis size is increased. RPA performs better, but our results do not suggest convergence with basis size. Because it is not clear if this is a fault of the approximation or of the basis for CI (which 
again is our numerically exact benchmark) it would be good, if and when possible, to continue this 
investigation with even larger bases.

\begin{table}

\caption[Electron Affinity]{Electron Affinity (hartree)\label{tab:affinity}}

\begin{tabular}{lr|rrrrrrrr|r}
\hline
\multicolumn{1}{c}{Atom} & \multicolumn{1}{c}{space} & \multicolumn{1}{c}{HF} &  & \multicolumn{1}{c}{PHF} & &\multicolumn{1}{c}{RPA} & & \multicolumn{1}{c}{CI} & & \multicolumn{1}{c}{Exp \cite{lide}}\\

\hline

  & CVB1 &-0.0152& &-0.0152& &0.0780& &0.0111& &  \\

Li  &CVB2  &0.0978 & &0.0050& &0.0804& &0.0226& &0.0227 \\ 

  &CVB3  &-0.0020& &0.0056& &0.0768& &0.0226& & \\ 

\hline


   & CVB1&-0.0379& &\multicolumn{1}{c}{0.0163}& &-0.0302& &-0.0141 & & \\

B   & CVB2 &-0.0222& &-0.0413& &0.0009& &0.0068 & &0.0102 \\

   &CVB3 &-0.0205& &-0.0004& &0.0455& &\multicolumn{1}{c}{--} & & \\

\hline

   & CVB1 &-0.0126& &0.0081& &0.0186& &0.0178& & \\

C   & CVB2 &0.0057 & &0.0203& &0.0591& &0.0446 & &0.0464 \\

   & CVB3 &0.0060& &0.0192& &0.0616& &\multicolumn{1}{c}{--}& &\\

\hline


   & CVB1 &-0.0110& &-0.0564& &0.0399& &-0.0047& &\\

O   & CVB2 &0.0089& &-0.0302& &0.1092& &\multicolumn{1}{c}{--}& &0.0537 \\

   & CVB3 &0.0148& &-0.0302& &0.1148& &\multicolumn{1}{c}{--}& & \\

\hline

   & CVB1 &0.0133 & &0.0090 & &0.0890& &0.0600& & \\

F   & CVB2&0.0437& &0.0405& &0.1706& &\multicolumn{1}{c}{--}& &0.1250 \\

   &CVB3 &0.0432& &0.0399& &0.1748&  &\multicolumn{1}{c}{--}& & \\

\hline
\end{tabular}
\end{table}

\subsection{Optimization of the space}

We utilized the CVB$n$ STO-basis sets , which had been arrived at 
through optimization  using both Hartree-Fock and CI calculations \cite{Ema03}. 
But CI calculations, particularly in large 
dimension spaces, are tedious and time-consuming.   Furthermore, a basis optimized for Hartree-Fock calculations may not be optimized for CI calculations. 
Therefore it might be useful to see how well our approximate methods can work as a 
proxy for full CI when changing the basis. In other words, if we vary the basis, 
does a minimum in CI have a corresponding minimum in one or more of our approximate 
methods?

\begin{figure}
\centering
\includegraphics[width=0.6\textwidth, trim=0mm 0mm 0mm -25mm]{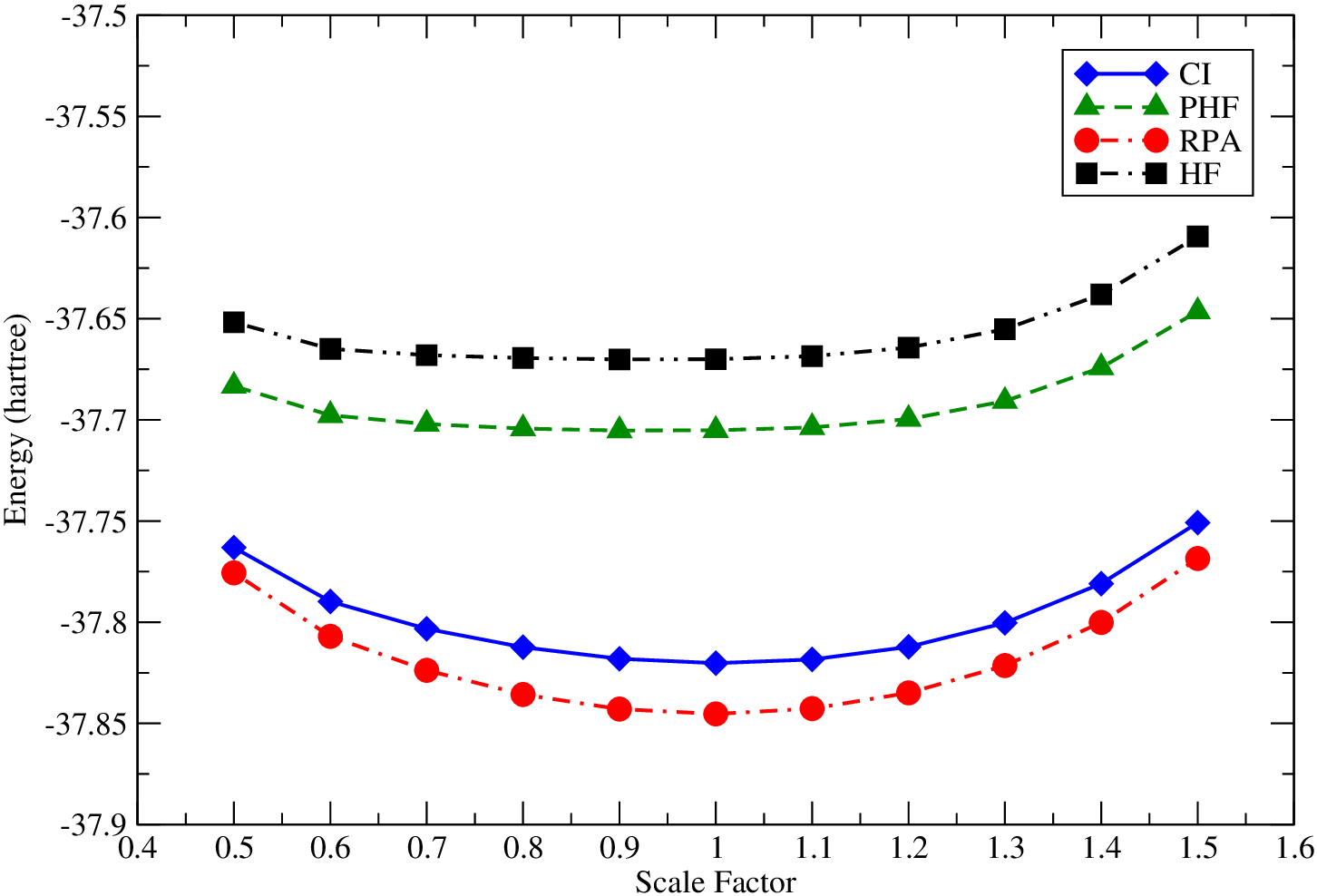}
\caption{(Color online.) Carbon optimization. As the STO basis parameters are scaled by a scale facter between $0.5$ and $1.5$, the random phase approximation tends to track CI very well. Hartree-Fock and projected Hartree-Fock follow each other but not CI. This indicates that RPA can be useful as a proxy for CI during basis parameter optimization. \label{fig:carbonopt}}
\end{figure}

We carried out an example optimization by altering all of the parameter values in the STO for carbon CVB1 by a scaling factor ranging from $0.5$ to $1.5$.
Table \ref{table:carbonminima} gives the results of a quadratic fit to each computational method.
The data show that RPA has a similar minimum to CI. Figure \ref{fig:carbonopt} shows these results graphically.
\begin{table}[hbt]
\centering
\caption{Carbon Minimum Energy Scale Factor \label{table:carbonminima}}
\begin{tabular}{lc}
\hline
 Method&Scale Factor\\
\hline
 CI &0.981\\
 RPA&0.989\\
 HF &0.892\\
 PHF&0.908\\
\hline
\end{tabular}
\end{table}

This optimization suggests that RPA is the most desirable method of those studied for basis set optimization. It is much faster than CI and clearly tracks the ground state energy as basis parameters are altered. PHF is a more useful upper bound to the ground state energy than HF, but cannot track the CI energy well enough to be used as a proxy for CI.


\section{Conclusions and Future work}

This paper is an extension of previous work comparing the random phase approximation against CI calculations 
in a CI basis \cite{SJ02}, testing \textit{how good is an approximation against a 
numerically exact result}.  By comparing against a numerically exact result, we 
eliminate any errors due to choice of model space, interaction, etc.  As part of this process 
we find eigenvalues of very large CI matrices. 
To simplify matters we used an uncoupled basis, which allowed us to recomputed the many-body matrix 
elements on the fly in a highly efficient manner. 

Not surprisingly, we found in absolute error RPA gave the best approximation to the 
ground state energy. RPA is not variationally bounded, and in the cases we chose it systematically overestimated the 
binding energy; this contrasts from previous, similar work for nuclei, where RPA both 
over- and underestimated the binding energy \cite{SJ02}.   Because the overestimate 
was not uniform, HF and PHF calculations sometimes  gave accidentally better estimates 
of the ionization potentials and electron affinities. 

Perhaps most usefully, we found that for an example fixed atomic system, RPA tracks the CI binding
as we varied parameters for the basis, suggesting RPA might provide an efficient 
proxy to speed up optimization of a basis.  This will be explored further in future work. 

In principle we can also compute and compare transitions and strength functions, 
as has been done for nuclei \cite{SJ03}, and static and dynamic 
observables such as static moments, polarizabilities, etc. This we also leave for future work. 

We thank Michael Bromley for many valuable conversations and suggestions, and in particular for aid 
in validating our codes. 
The U.S.~Department of Energy supported this investigation through
grants DE-FG02-96ER40985 and DE-FC02-07ER41457.


\begin{thebibliography}{99}

\bibitem{SJ02}
I. Stetcu and C.W. Johnson, Phys. Rev. C \textbf{66}, 034301 (2002).

\bibitem{SJ03}
I. Stetcu and C.W. Johnson, Phys. Rev. C \textbf{67}, 044315 (2003);
I. Stetcu and C.W. Johnson, Phys. Rev. C \textbf{69}, 024311 (2004).

\bibitem{PHF} J. T. Staker and C. W. Johnson, to be published.


\bibitem{LM85} I. Lindgren and J. Morrison, \textit{Atomic many-body theory}, 2nd. ed. (Springer-Verlag, Berlin, 1985).  

\bibitem{BG77} P.J. Brussard and P.W.M. Glaudemans, \textit{Shell-model applications 
in nuclear spectroscopy} (North-Holland Publishing Company, Amsterdam 1977). 

\bibitem{ring}
P. Ring and P. Shuck, \textit{The nuclear many-body problem}, 1st edition,
Springer-Verlag, New York 1980.


\bibitem{Ema03} I.~Ema \textit{et al.}, J. Comput. Chem. \textbf{24}, 859 (2003).

\bibitem{Cook98} D. B. Cook, \textit{Handbook of computational quantum chemistry,}
(Oxford University Press, Oxford, 1998).

\bibitem{FFBJ97} C. Froese Fischer, T. Brage and P. J\"onsson, \textit{Computational atomic structure:
an MCHF approach} (Institute of Physics Publishing, Bristol, 1997).


\bibitem{Je07} F. Jensen, \textit{Introduction to computational chemistry,} 2nd ed. (John Wiley 
$\&$ Sons Tld, West Sussex, 2007).


\bibitem{Lo55CI} P.-O. L\"owdin, Phys. Rev. \textbf{97}, 1474 (1955) 


\bibitem{SS99}  C.D. Sherrill and H.F. Schaefer III, Adv. Quant. Chem. \textbf{34}, 143
(1999).



\bibitem{BW88}  B. A. Brown and B. H. Wildenthal, Annu. Rev. Nucl. Part. Sci. 38, 29 (1988).

\bibitem{Sh98} I. Shavitt, Mol. Phys. \textbf{94}, 3 (1998).


\bibitem{CIV3} A. Hibbert, Comp. Phys. Comm. \textbf{3}, 
141 (1975); A. Hibbert, Comp. Phys. Comm. \textbf{35}, 
C-298 (1984).

\bibitem{GRASP92}  F.A. Parpia, C. Froese Fischer, and I. P. Grant, Comp. Phys. Comm. \textbf{94}, 
249 (1996);  F.A. Parpia, C. Froese Fischer, and I. P. Grant, Comp. Phys. Comm. \textbf{175}, 
745 (2006).

\bibitem{BIGSTICK} W. E. Ormand and C. W. Johnson, to be published.


\bibitem{ANTOINE} E.~Caurier and F.~Nowacki, Acta Phys. Pol.
 B \textbf{30} (1999) 705 


\bibitem{Deb09} N. C. Deb and A. Hibbert, At. Data and Nucl. Data Tables, \textbf{95}, 184 (2009).



\bibitem{Correge04} G. Corr\'eg\'e and A. Hibbert, At. Data and Nucl. Data Tables, \textbf{86}, 19 (2004).

\bibitem{Gupta05} G. P. Gupta and A. Z. Msezane, At. Data and Nucl. Data Tables, \textbf{89}, 1 (2005).

\bibitem{Chen01} M. H. Chen, K. T. Cheng, and W. R. Johnson, Phys. Rev. A \textbf{64}, 042507 (2001).

\bibitem{Bro10} M. W. J. Bromley and J. Mitroy, Phys. Rev. A \textbf{81}, 052708 (2010).

\bibitem{OJS90} J. Olsen, P. J{\o}rgensen, and J. Simons, Chem. Phys.Lett. \textbf{169}, 463 (1990). 

\bibitem{Lanczos} R. R. Whitehead, A. Watt, B. J. Cole, and I. Morrison, Advances in Nuclear Physics (Plenum, New York, 1977), Vol. 9, p. 123.
\bibitem{Dav93} E. R. Davidson, Comput. Phys. \textbf{7}, 519 (1993). 

\bibitem{LSAS01} M. L. Leininger, C. D. Sherrill, W. D. Allen, and H. F. Schaefer III, 
J. Comp. Chem. \textbf{22}, 1574 (2001). 


\bibitem{Weiss61} A. W. Weiss, Phys. Rev. \textbf{122} (6), 1826-1836 (1961).

\bibitem{Bu68} C. F. Bunge, Phys. Rev. \textbf{168}, 92 (1968) 



\bibitem{Saski74} F. Saskai and M. Yoshimine, Phys. Rev. A \textbf{9} (1), 17-25 (1974).


\bibitem{Bunge76} C. F. Bunge, Phys. Rev. A \textbf{14} , 1965 (1976);  At. Data Nucl. Data Tables 
\textbf{18}, 126 (1976). 

\bibitem{Silver79} D. Silver, S. Wilson, and C. F. Bunge, Phys. Rev. A \textbf{19} (4), 1375-1382 (1975).


\bibitem{Jitrik97} O. Jitrik and C. F. Bunge, Phys. Rev. A \textbf{56} (4), 2614-2623 (1997).

\bibitem{Bunge06} C. F. Bunge, J. Chem. Phys. \textbf{125}, 014107 (2006).


\bibitem{Almora10} C. X. Almora-Diaz, and C. F. Bunge, Int. J. Quantum Chem. \textbf{110} (15), 2982-2988 (2010).

\bibitem{Lo55HF} P.-O. L\"owdin, Phys. Rev. \textbf{97}, 1490 (1955) 


\bibitem{Lunell68} S. Lunell, Phys. Rev.  \textbf{173}, 86 (1968).

\bibitem{Lask75} B. Laskowski and S. Lunell, Int. J. Quantum Chem. \textbf{9} (S9), 175-182 (1975).



\bibitem{La80} R. D. Lawson, \textit{Theory of the nuclear shell model} (Clarendon Press, 
Oxford, 1980).

\bibitem{Edmonds} A. R. Edmonds, \textit{Angular momentum in quantum mechanics} (Princeton University Press, New Jersey, 1996).

\bibitem{Lang93} G. H. Lang, C. W. Johnson, S. E. Koonin,  and W. E. Ormand, Phys. Rev. C  
\textbf{48}, 1518 (1993);
E. Y. Loh, Jr. and J. E. Gubernatis, in \textit{Electronic Phase Transitions}, edited by W. Hanke and Yu. V. Kopaev (Elsevier Science Publishers B.V., New York, 1992).



\bibitem{lide} David R. Lide, \textit{CRC Handbook of Chemistry and Physics} 85th edition, (CRC Press Inc, Florida 2005).

\bibitem{SHERPA} I. Stetcu, PhD. thesis, Louisiana State University 2003 (unpublished).





\bibitem{Hotop75} H. Hotop and W. C. Lineberger, J. Phys. Chem. Ref. Data \textbf{4} (3), 539-576 (1975).




\bibitem{Jiang07} H. Jiang and E. Engel, J. Chem. Phys. \textbf{127}, 184108 (2007).

\bibitem{Gruneis09} A. Gr\"{u}neis \textit{et al.}, J. Chem. Phys. \textbf{131}, 154115 (2009).

\bibitem{Cole82} L. A. Cole and J.  F. Perdew, Phys Rev A \textbf{25}, 1265 (1982).





\end{thebibliography}
\end{document}